\newcommand{\shorttitle}{Characterization of the State of Hydrogen} 
\newcommand{\shortauthor}{B. Militzer, W. Magro, D. Ceperley} 
\newcommand{\contnum}{} 
\newcommand{\contvolum}{39} 
\newcommand{\contyear}{1999} 
\newcommand{\contpages}{151-154} 
\newcommand{\HH}{{\mathcal{H}}}
\newcommand{\PP}{{\mathcal{P}}}
\newcommand{\RR}{{\mathcal{R}}}
\documentstyle[twoside,epsfig,amssymb]{cpp}
\begin{document}
\setcounter{page}{101} 
\title{Characterization of the State of Hydrogen at High Temperature and Density} 
\author{Burkhard Militzer(a), William Magro(b), David Ceperley(a)} 
\address{(a) National Center for Supercomputing Applications, Department of Physics, 
University of Illinois, Urbana, IL 61801\\
(b) William Magro, Kuck \& Associates, Inc., Champaign, IL 61820} 
\maketitle
\begin{abstract}   
Fermionic path integral Monte Carlo simulations have been applied
to study the equilibrium properties of the hydrogen and deuterium 
in the density and temperature range of $1.6 < r_s < 14.0$ and 
$5000K < T < 167000K$.  We use this technique to determine the 
phase diagram by identifying the plasma, the molecular, atomic and metallic regime.
We explain how one can identify the phases in the path
integral formalism and discuss the state of hydrogen for 5 points in the
temperature-density plane. 
Further we will provide arguments for the nature of the transitions 
between the regimes.
\end{abstract}
%
%
\section{Introduction}
The phase diagram of hydrogen has been studied intensively with different 
theoretical approaches \cite{rotesbuch},\cite{Sa92}, simulation techniques 
\cite{Le97},\cite{Na98} and experiments \cite{Da97},\cite{WeMi96}.
From theory, the principal effects at low densities are
well-known. On the other hand, the properties at intermediate density are not yet 
well understood, and the phase diagram is not yet accurately
determined. In particular, the nature of the transition to a metallic
state is still an open question.

\begin{figure}[h] 
\begin{center}
\epsfig{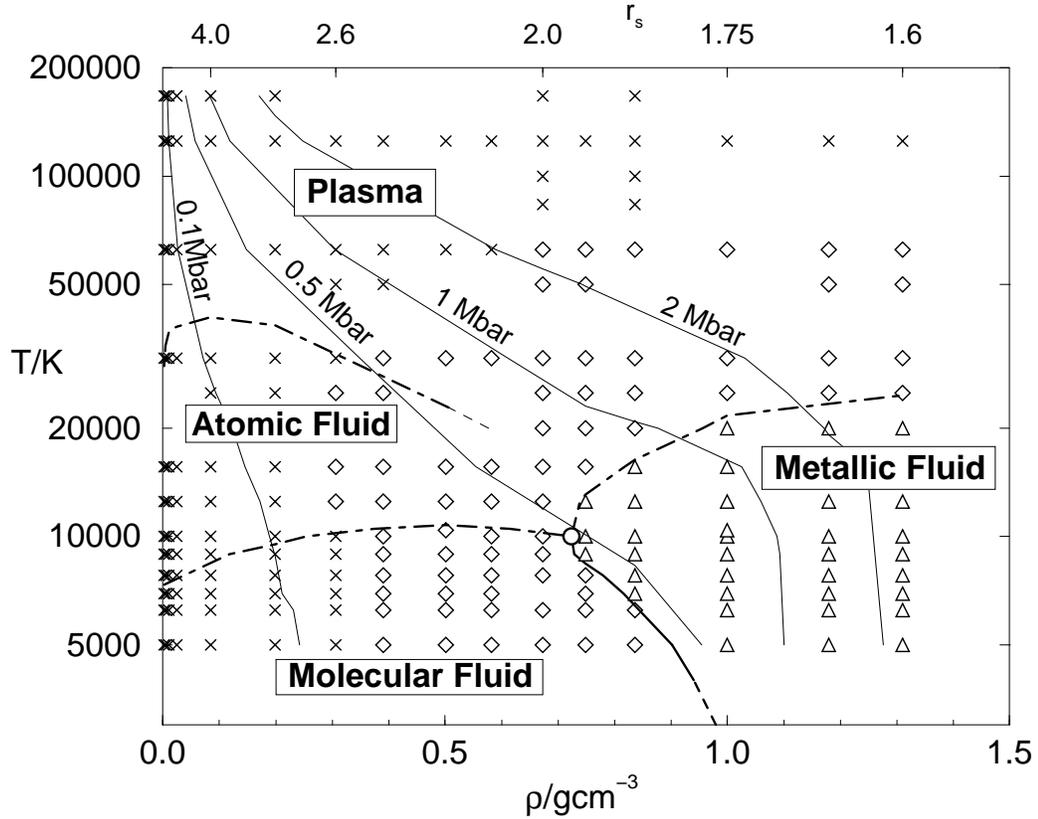}
\end{center}
\caption{The computed phase diagram of deuterium is shown in the 
temperature-density plane. $(\times,\diamond,\bigtriangleup)$ indicate our PIMC 
simulations and distinguish between different degrees of degeneracy of the 
electrons ($\times$ less than 10\% exchanges, $\diamond$ more 10\% and 
$\bigtriangleup$ over 80\%). 
The four main regimes, molecular, atomic and metallic fluid as well as the
plasma are shown. The thick solid line specifies the plasma phase transition predicted
in \cite{Ma96}. The thin solid lines specify the approximate location of isobars.}
\label{fig1}
\end{figure}

In this article, we would like to show how these questions can be addressed
by path integral Monte Carlo (PIMC) simulations. Using this
approach, we derived the phase diagram in Fig.\ref{fig1} where we
distinguish between molecular, atomic, metallic and plasma regimes. We will
demonstrate how these different states can be identified from PIMC simulations.
\begin{figure}[p] 
\begin{center}
\epsfig{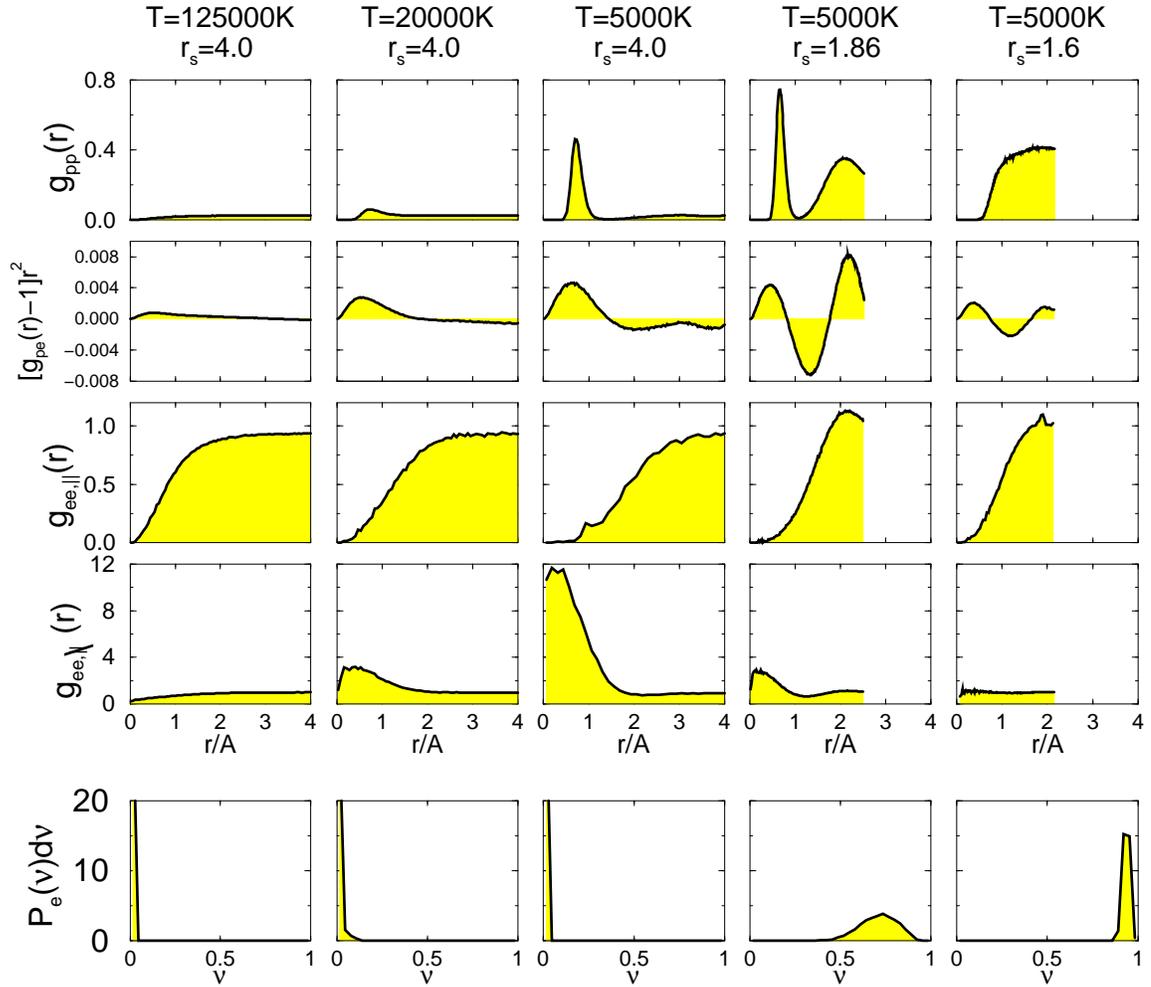}
\end{center}
\caption{Distribution functions for a selection of 5 simulation of hydrogen at different temperatures
and densities, one in each column: 1) a plasma, 2) a atomic fluid, 3) a molecular fluid, 4)
a molecular fluid with metallic properties, and 5) a metallic fluid. 
The rows show the following: (1) proton-proton correlation function
$g_{pp}(r)$ multiplied by the density, which means the area under the peak at the bond length of $r=0.75$\AA~
 indicates the number of molecules, (2) $[g_{pe}(r)-1]r^2$, where the first peak hints to 
the existence of bound electrons in the ground state, (3) pair correlation function for electrons
with parallel spins demonstrating the Pauli exclusion principle, 
(4)  pair correlation function for electrons with anti-parallel spins, where the peak is caused
by a localization of wave function along the molecular bond, and
(5) distribution of the fraction $\nu$ of electrons involved in a permutation. A peak near $\nu=0$
represent a small degree of degeneracy of the electrons, while one near $\nu=1$ implies a highly
degenerate electron gas. }
\label{fig2}
\end{figure}
The imaginary-time path integral formalism \cite{Ce95} is based on the position-space 
density matrix $\rho(\RR, \RR', \beta)$, which can be used to determine
the equilibrium expectation value of any operator $\hat{O}$,
\begin{equation}
\langle \hat{O} \rangle = \frac{\mbox{Tr} \, \hat{O} \rho}{\mbox{Tr}\, \rho }
=\frac{ \int d\RR d\RR' \, \rho(\RR, \RR', \beta) \, 
\langle \RR | \hat{O} | \RR' \rangle }
{ \int d\RR \, { \rho(\RR,\RR, \beta)} }
\end{equation}
where $\RR$ represents the coordinates of all particles. The low temperature
density matrix $\rho(\RR,\RR', \beta) = \langle \RR | e^{-\beta \HH } | \RR' \rangle $ 
can be expressed as product of high
temperature density matrices  $\rho(\RR,\RR, \tau)$ with the time step $\tau = \beta /M$.
In position space, this is a convolution,
\begin{equation}
\label{conv}
\rho(\RR_0,\RR_M;\beta) = \int\!\cdots\!\int d\RR_{1}\:d\RR_{2}\:\cdots\: d\RR_{M-1}\;
\rho(\RR_0,\RR_{1}; \tau ) \: \rho(\RR_{1},\RR_{2};\tau ) \: \cdots \:
\rho(\RR_{M-1},\RR_M ;\tau ). 
\end{equation}
This high dimensional integral can be integrated using Monte Carlo methods. Each
particle is represented by a closed path in imaginary time.
Fermi statistics is taken into account by considering the fermion density
matrix, which can be expressed by considering all permutations $\PP$ of identical
particles,
\begin{equation}
\rho_F(\RR,\RR';\beta) = {\mathcal{A}} \rho(\RR,\RR';\beta) = 
\frac{1}{N!}\sum_{\PP} (-1)^\PP \rho(\RR,\PP \RR';\beta),
\end{equation}
where $\mathcal{A}$ is the antisymmetrization projection operator. Cancellation
of positive and negative contributions leads to the {\sl fermion sign problem},
which is solved approximately by restricting the paths within a nodal surface derived
from the free-particle density matrix \cite{Ce96}.
\section{Phase diagram of hydrogen and deuterium}

We used PIMC simulation with 32 protons and 32 electrons and a time step $\tau=1/10^6\,$K
to generate the phase diagram shown in Fig. \ref{fig1}.
In the low density and low temperature regime, we find a  
molecular fluid. In the proton-proton correlation function shown in Fig.
\ref{fig2}, one finds a clear peak at the bond length of $0.75\,$\AA. We determine
the number of molecules as well as other compound particles by a cluster analysis based
on the distances. Using this approach we can estimate the number of
bound states (see \cite{Mi97}). We can also estimate the fraction of molecules
and atoms to determine the regime boundaries. However
at high density, a clear definition of those species is difficult to give.

Starting in the molecular regime, one finds that increasing temperature at constant density
leads to gradual dissociation of molecules followed by a regime, with a majority of atoms.
The atoms are then gradually ionized at 
even higher temperatures. Lowering the density at constant temperature 
leads to a decrease in the number of molecules, or atoms respectively, 
due to entropy effects. 

If the density is increased at constant temperature, pressure dissociation diminishes the
molecular fraction. This transition was described by Magro et. al. \cite{Ma96}. 
Its precise nature is still a topic of our current research. Using PIMC simulations, 
one finds it occurs within a small density interval and 
we predict that it is connected with both the molecular-atomic and insulator-metal
transition.
We determine the fraction of electrons involved
in a permutation as an indication of electronic delocalization. 
Permuting electron are required to form a Fermi surface, which
means that a high number of permutations indicate a high degree of degeneracy of
the electrons. Permuting electrons form long chains of paths and therefore occupy
delocalized states. This delocalization destabilizes the hydrogen molecules. 
Before all bonds are broken,
one finds a molecular fluid with some permuting electrons, which could indicate
the existence of a molecular fluid with metallic properties.

The boundaries of the metallic regime are determined by two effects. With 
increasing temperature, the degree of degeneracy of the electrons is simply reduced.
If the temperature is lowered, the attraction to the protons becomes more relevant,
which localizes the electron wave function and decreases the degree of degeneracy also
(see Fig. \ref{fig1}).
\begin{acknowledgements}
Support from CSAR program and computer facilities at NCSA and Lawrence Livermore National Laboratory.
\end{acknowledgements}
\begin{received}Received October 1, 1998
\end{received}
\end{document}